\documentclass[showpacs,aip,apl,amsmath,amssymb,reprint]{revtex4-1}
\usepackage{graphicx}
\usepackage{bm}
\usepackage{color}

\begin{document}

\title{Current-driven thermo-magnetic switching in magnetic tunnel junctions}

\author{A.~F.~Kravets}
\email{anatolii@kth.se}
\author{D.~M.~Polishchuk}
\affiliation{Nanostructure Physics, Royal Institute of Technology, 10691 Stockholm, Sweden}
\affiliation{Institute of Magnetism, National Academy of Sciences of Ukraine, 03680 Kyiv, Ukraine}

\author{V.~A.~Pashchenko}
\affiliation{B. Verkin Institute for Low Temperature Physics and Engineering, National Academy of Science of Ukraine, 61103 Kharkiv, Ukraine}

\author{A.~I.~Tovstolytkin}
\affiliation{Institute of Magnetism, National Academy of Sciences of Ukraine, 03680 Kyiv, Ukraine}

\author{V.~Korenivski}
\affiliation{Nanostructure Physics, Royal Institute of Technology, 10691 Stockholm, Sweden}

\begin{abstract}
We investigate switching of magnetic tunnel junctions (MTJs) driven by the thermal effect of the transport current through the junctions. The switching occurs in a specially designed composite free layer, which acts as one of the MTJ electrodes, and is due to a current-driven ferro-to-paramagnetic Curie transition with the associated exchange decoupling within the free layer leading to magnetic reversal. We simulate the current and heat propagation through the device and show how heat focusing can be used to improve the power efficiency. The Curie-switch MTJ demonstrated in this work has the advantage of being highly tunable in terms of its operating temperature range, conveniently to or just above room temperature, which can be of technological significance and competitive with the known switching methods using spin-transfer torques.
\end{abstract}

\maketitle
\date{\today}

Spin-dependent effects inherent to nanosized magnetic systems, such as giant,\cite{Baibich1988,Binasch1989} and tunneling magnetoresistance,\cite{Moodera1995,Miyazaki1995} spin-transfer torques,\cite{Slonczewski1996,Berger1996,Brataas2012} etc., form the base for spin-electronic applications. Adding a new degree of freedom through thermal control broadens the physics as well as the functionality of the spin-thermo-electronic devices. Temperature is used for controlling the antiferromagnetic exchange pinning\cite{Prejbeanu2004} and switching in the commercially available thermo-assisted  magnetoresistive random-access memory.\cite{Prejbeanu2007,Prejbeanu2013} Thermal control has been proposed for on/off switching of the interlayer exchange coupling,\cite{Thiele2008,Andersson2010} relevant for storage media or spin-thermionic oscillators\cite{Kadigrobov2010} as well as for spin-transfer torque in nano-devices pumped by thermal magnons.\cite{Slonczewski2010} Implementation of these effects on the nanoscale, with the accompanying requirements of high-scale device integration, necessitates an intrinsic heating mechanism -- typically Joule heating in suitable conductive elements of the device. A demonstration of new functionality ideally is done using a fully integrated device, incorporating the heating and switching elements, with the magnetic switching element ideally being the widely used magnetic tunnel junction (MTJ).

\begin{figure*}
\includegraphics[width=15cm]{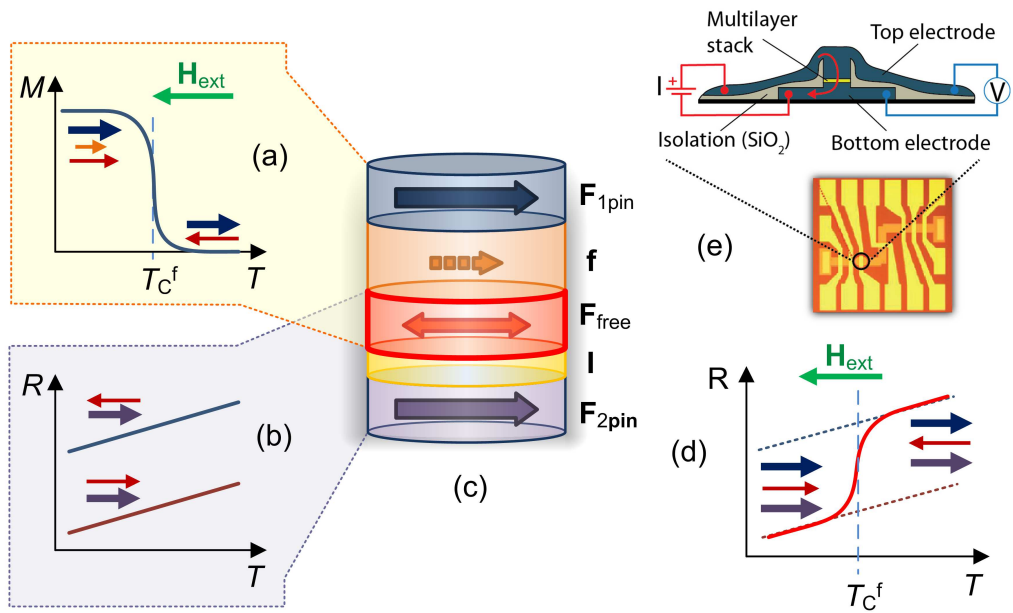}
\caption{Illustration of thermo-magnetic switching in a Curie-valve F$_{\textrm{1pin}}$/f/F$_{\textrm{free}}$ integrated with magnetic tunnel junction F$_{\textrm{free}}$/I/F$_{\textrm{2pin}}$ into a vertical stack: (a)~total magnetic moment $M$~vs.~temperature $T$ for the Curie-valve; (b)~resistance $R$~vs.~$T$ for parallel and antiparallel states of the MTJ  in the absence of thermo-magnetic switching; (c)~layout of the stack with shared free layer F$_{\textrm{free}}$ between the Curie-valve and the MTJ; (d) $R(T)$ for the Curie-valve-MTJ structure with thermo-magnetic switching; (e)~side (illustration) and top (optical microscope image) of the Curie-valve-MTJ multilayered stack fabricated using three-step lithography. Arrows in (a)--(d) indicate orientation of magnetic moments of individual layers.}\label{figure1}
\end{figure*}

The temperature-dependent interlayer exchange coupling in the so-called Curie-valves of the generic structure F$_{\textrm{free}}$/f/F$_{\textrm{pin}}$ has been studied in some detail.\cite{Andersson2010,Kravets2012,Kravets2016} Here, in contrast to the conventional spin-valves of structure F$_{\textrm{free}}$/N/F$_{\textrm{pin}}$, the nonmagnetic spacer (N) between the free (F$_{\textrm{free}}$) and exchange-pinned (F$_{\textrm{pin}}$) strongly ferromagnetic (FM) layers is replaced with a weakly ferromagnetic layer (f) having a low Curie temperature ($T_{\textrm{C}}^{\textrm{f}}$). Below $T_{\textrm{C}}^{\textrm{f}}$ the spacer f is in its ferromagnetic state and the two outer ferromagnetic layers F$_{\textrm{free}}$ and F$_{\textrm{pin}}$ are exchange coupled through the spacer.\cite{Kravets2015} Above $T_{\textrm{C}}^{\textrm{f}}$ the outer ferromagnetic layers are essentially decoupled. As a result, with a relatively weak external magnetic field applied antiparallel to the exchange-bias direction of F$_{\textrm{pin}}$, a change in temperature across the Curie point of the spacer f induces switching from the saturation magnetic moment of the valve to zero, illustrated in Fig.~\ref{figure1}(a). A dilute ferromagnetic alloy of variable concentration used as the spacer material offers simplicity and flexibility in controlling the key thermo-magnetic switching parameters, such as the temperature and transition width, with the operating point ideally just above room temperature.\cite{Kravets2014}

To provide such intrinsic thermal control we have integrated a Curie-valve F$_{\textrm{1pin}}$/f/F$_{\textrm{free}}$ with a magnetic tunnel junction F$_{\textrm{free}}$/I/F$_{\textrm{2pin}}$, where F$_{\textrm{free}}$ is the switching as well as the readout layer, I is the insulating tunnel barrier [Fig.~\ref{figure1}(c)]. Low-resistance MTJs besides their useful magnetoresistance were shown to efficiently generate localized heating due to inelastic relaxation of tunneling electrons in the immediate vicinity of the barrier.\cite{Sousa2004,Akerman2006,Oliver2004} Thus constructed Curie-valve-MTJ stacks should then exhibit thermo-magnetic switching in a weak magnetic field, schematically shown in Fig.~\ref{figure1}(d). Below we experimentally demonstrate such current-induced magnetic switching in a vertical MTJ pillar with the current-perpendicular-to-the-plane (CPP) [Fig.~\ref{figure1}(e)] and perform a comprehensive analysis of the relevant magneto-thermo-electronic properties.

Multilayers of general layout bottom-electrode/[F$_{\textrm{2pin}}$/I/F$_{\textrm{free}}$/f/F$_{\textrm{1pin}}$]/top-electrode were deposited on thermally oxidized Si substrates using magnetron sputtering. Here, F$_{\textrm{2pin}}$ = IrMn(15)/CoFe(2)/CoFeB(2), I = AlO$_x$(0.7), F$_{\textrm{free}}$ = CoFeB(5), f = Ni-Cu(20), F$_{\textrm{1pin}}$ = CoFe(2)/IrMn(12), with all thicknesses given in parenthesis in nm. The bottom and top electrodes were thick Cu layers (up to 100~nm in thickness), with Ta(5) seed layers between the Cu electrodes and the inner multilayer stack. Antiferromagnetic Ir$_{20}$Mn$_{80}$ (IrMn) was used to exchange bias the Co$_{90}$Fe$_{10}$ (CoFe) reference layers of F$_{\textrm{2pin}}$ and F$_{\textrm{1pin}}$. Low-coercivity Co$_{60}$Fe$_{20}$B$_{20}$ [CoFeB(5)] layer is used to perform the role of the shared free layer F$_{\textrm{free}}$. The Ni$_{72}$Cu$_{28}$ [Ni-Cu(20)] layer was co-sputtered from individual Ni and Cu targets. In order to induce exchange bias at the IrMn/FM interfaces, the respective deposition was performed in a magnetic field of $\sim$600~Oe. Magnetization measurements of reference (un-patterned) multilayers were performed using a SQUID magnetometer MPMS-XL5 Quantum Design, in the temperature range of 300--370~K, and in magnetic fields up to 1~kOe applied in the film plane and parallel to the pinning direction of F$_{\textrm{1pin}}$ and F$_{\textrm{2pin}}$. A three-step photolithography process was used for fabrication of the F$_{\textrm{2pin}}$/I/F$_{\textrm{free}}$/f/F$_{\textrm{1pin}}$ junctions of 3~$\mu$m in lateral diameter, patterned at the center of crossing 50~$\mu$m wide bottom and top electrodes [Fig.~\ref{figure1}(e)]. CPP magnetoresistance measurements were performed using the four-point method, with the magnetic field applied in-plane along the pinning direction.

\begin{figure*}
\includegraphics[width=13cm]{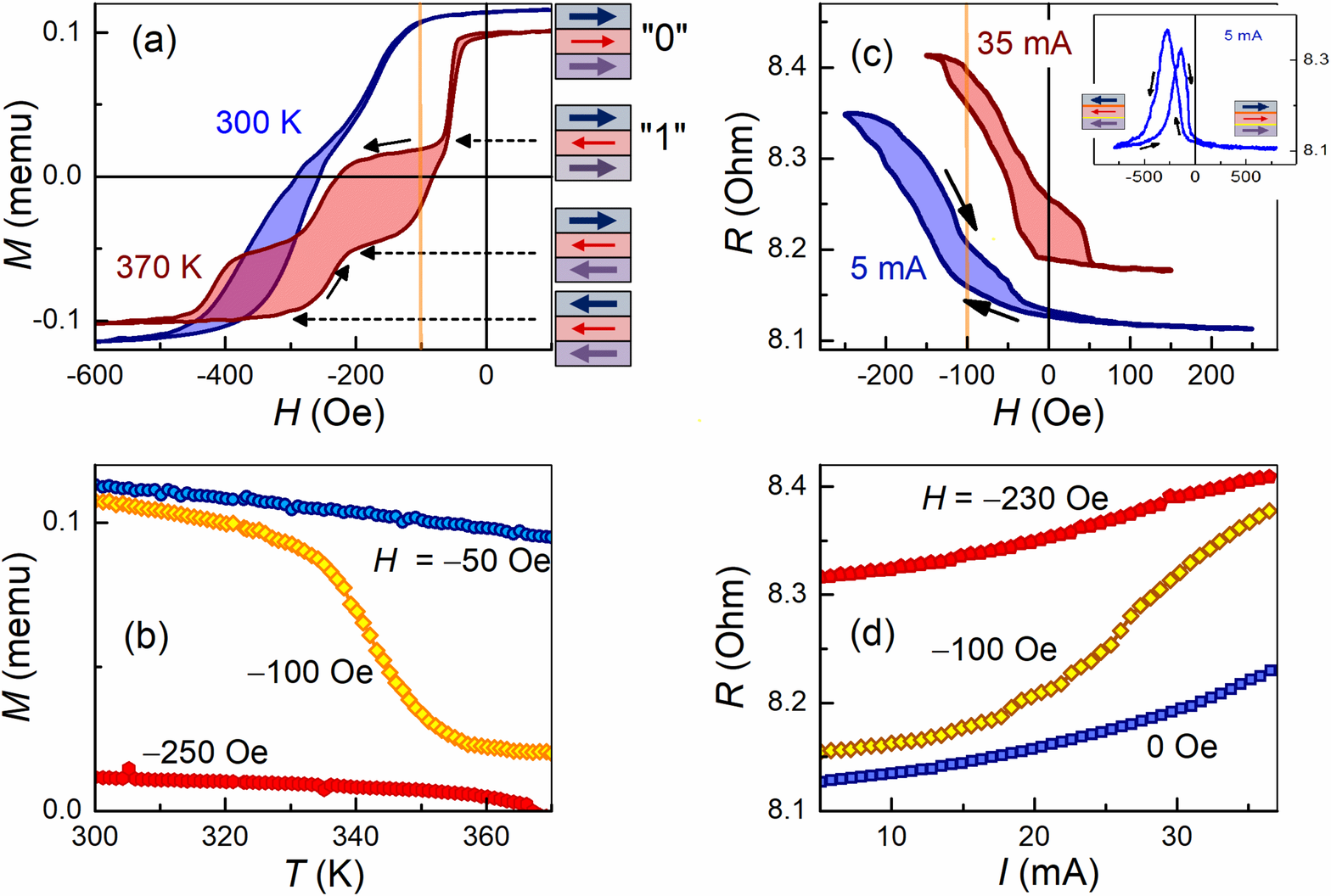}
\caption{(a) Hysteresis loops $M(H)$ of reference thin-film multilayer at 300 and 370~K, measured along exchange-pinning direction. (b) $M(T)$ for three fixed magnetic field values. (c) Minor loops of resistance vs. magnetic field, $R(H)$, for patterned MTJ measured at two representative values of transport current (CPP MTJ configuration). Inset shows major $R(H)$ loop recorded at 5 mA. (d) Resistance as a function of current measured at different values of applied magnetic field.}\label{figure2}
\end{figure*}

A magnetometric study of the reference samples was used to obtain information on the key characteristics of their thermomagnetic behavior, summarized in Figs.~\ref{figure2}(a) and \ref{figure2}(b). The magnetization loop $M(H)$, being typical of an exchange coupled multilayer at 300~K, becomes substantially transformed on increasing the temperature above room temperature (RT), eventually becoming characteristic of a decoupled, spin-valve-type structure at 370~K [Fig.~\ref{figure2}(a)]. Such temperature evolution of $M(H)$ is due to a ferromagnetic-to-paramagnetic transition in the Ni-Cu layer used as the low-$T_{\textrm{C}}$ spacer. At $T<T_{\textrm{C}}^{\textrm{f}}$ the spacer is ferromagnetic and exchange couples the layers F$_{\textrm{free}}$ and F$_{\textrm{1pin}}$, which then behave as one exchange-pinned ferromagnetic layer. This is seen as a one merged $M(H)$ transition at 300~K, field-offset by exchange-pinning. At $T>T_{\textrm{C}}^{\textrm{f}}$ the spacer f is paramagnetic and made sufficiently thick not to transit magnetic exchange between F$_{\textrm{free}}$ and F$_{\textrm{1pin}}$. The resulting $M(H)$ loop at 370~K is composed of three minor loops, which reflect the magnetization reversal in the three individual FM layers. Tracing the positive-to-negative field sweep, one is able to put minor loops in relation to the magnetization orientation in the FM layers. At $H>0$, all magnetic moments of F$_{\textrm{free}}$, F$_{\textrm{1pin}}$ and F$_{\textrm{2pin}}$ are aligned along the pinning direction (state ``0''). On application of a negative field ($H<0$), the F$_{\textrm{free}}$ layer switches first, at $H \approx -70$~Oe (state ``1''). Next, the F$_{\textrm{2pin}}$ and F$_{\textrm{1pin}}$ layers switch at $H \approx – 280$ and $– 450$~Oe, respectively. These latter switching fields are different due to the exchange-pinning of F$_{\textrm{2pin}}$ by the adjacent antiferromagnetic IrMn.

As indicated in Fig.~\ref{figure2}(a) by the yellow vertical line, it is possible to switch between state ``0'' and state ``1'' by changing temperature only, in this case from 300~K to 370~K. The $M(T)$ dependence for $H = - 100$~Oe in Fig.~\ref{figure2}(b) demonstrates such thermally-induced switching of the free layer, between the parallel and antiparallel states of the Curie-valve as well as the MTJ. Shown also is that applied fields of 0~Oe or $-250$~Oe result in one, parallel ground state in the full operating temperature range (here 300--370~K).

CPP magnetotransport measurements were performed on patterned vertical-pillar devices. The measurements showed a pronounced dependence of the electrical resistance $R$ on the applied magnetic field $H$ [Figs.~\ref{figure2}(c) and \ref{figure2}(d)]. $R(H)$ closely follow the $M(H)$ characteristics obtained on the respective un-patterned multilayers, indicating that the revealed magneto-resistive states are determined by the mutual orientation of the magnetic moments defining the MTJ, \textbf{M}$_{\textrm{free}}$ and \textbf{M}$_{\textrm{2pin}}$. A comparison of the $R(H)$ loop of Fig.~\ref{figure2}(c) (inset) and the corresponding 300~K $M(H)$ loop of Fig.~\ref{figure2}(a) supports this conclusion: two saturated states with \textbf{M}$_{\textrm{free}}$ and \textbf{M}$_{\textrm{2pin}}$ parallel yield lower resistance, whereas higher resistance is due to a misalignment of \textbf{M}$_{\textrm{free}}$ and \textbf{M}$_{\textrm{2pin}}$. 

The shape of $R(H)$ changes significantly on driving electrical current through the MTJ of sufficient magnitude. With increasing the current, the transition to the high resistance state occurs at lower fields [Fig.~\ref{figure2}(c)]. In a close correlation with the variable-temperature $M(H)$ behavior, $R(H)$ for 5~mA and 35~mA correspond to two magnetic states of the MTJ with switched F$_{\textrm{free}}$, between its exchange-coupled and exchange-decoupled configurations within the Curie-valve (through spacer f to layer F$_{\textrm{1pin}}$). For our micrometer scale junctions, with the effective Curie point of the spacer $T_{\textrm{C}}^{\textrm{f}}=350$~K without any specific heat focusing, current densities of the order of 1~mA/$\mu$m$^{2}$ are sufficient for switching the MTJ's state.

Spin-thermo-electronic switching is demonstrated in $R$~vs.~$I$, measured with suitable weak magnetic biasing [Fig.~\ref{figure2}(d)]. For $H=0$ and $-230$ Oe, $R(I)$ show the low and high resistive branches, respectively, whereas for $H = - 100$~Oe, a clear transition occurs between these two resistive states, driven by the effect of the Joule heating from the transport current. The $R$--$T$ transition observed is well controlled and fully reversible in the current range shown. Signs of junction degradation were noticed at currents in access of 50~mA. 
	
\begin{figure}
\includegraphics[width=8cm]{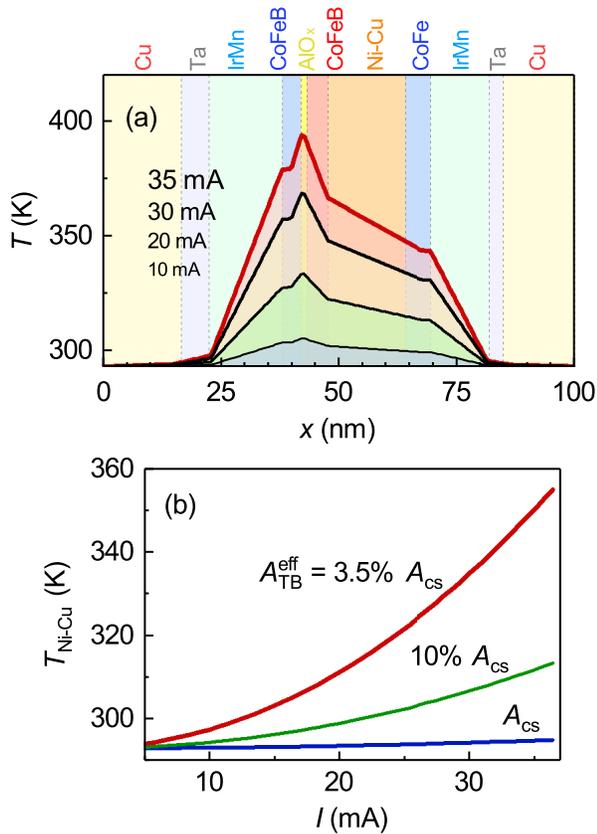}
\caption{(a)~Calculated temperature profiles for multilayer under electric current $I = 10$, 20, 30 and 35~mA, under assumption that effective tunnel-barrier area is 3.5\% of actual (see text for details). (b) Average temperature of Ni-Cu spacer layer, $T_{\textrm{Ni-Cu}}$, versus current, $I$, obtained numerically using Eq.~(\ref{equation1}) for effective junction area of $A^{\textrm{eff}}_{\textrm{TB}} = 100$\%, 10\% and 3.5\% of actual cross-section area, $A_{\textrm{cs}}$ (3~$\mu$m diameter).}
\label{figure3}
\end{figure}

To obtain more insight into the processes behind the observed current-induced thermo-magnetic switching, we have performed numerical simulations of the temperature distribution along the normal to the multilayer using the heat equation, in one dimension appropriate in our case. It has already been shown that numerical solutions to the equation obtained using the finite-element method are consistent with the experimental results for structures containing tunnel barriers.\cite{Sousa2004} The heat equation for our case has the following form:

\begin{equation}\label{equation1}
c_{p} d \dfrac{\partial T}{\partial t} - K \dfrac{\partial^{2} T }{\partial x^{2}} = F(j,x),
\end{equation}
where $c_{p}$ is the heat capacity, $d$ is the mass density, $K$ is the thermal conductivity of the material, $T$ is the temperature, and $t$ is the time. Function $F(j,x)$ is responsible for the heat generation when the electrical current of density $j$ passes through a metallic layer ($F = \rho j^{2}$), or through a tunnel barrier [$F= (jV/l)\exp( - x/l)$]. Here, $\rho$ is the electrical resistivity, $x$ is the stack position, $l$ is the inelastic scattering mean free path, and $V$ is the bias voltage. The physical parameters of the materials used in the simulation are listed in Table B1 of supplementary material. The results of the simulations presented in Fig.~\ref{figure3} and discussed below necessarily are of approximate character, but quite illuminating for the qualitative analysis of the switching mechanism demonstrated. 

The initial calculations were carried out using the junction cross section corresponding to a disc of 3~$\mu$m in diameter. The calculated Joule heating efficiency for the highest current in the simulated range (37 mA) is rather low, $\sim 0.7$~K/(mW $\mu$m$^{-2}$), resulting in the temperature rise of the Ni-Cu spacer layer of a few K [Fig.~\ref{figure3}(b)]. Similar estimates and experimental parameters were reported for micrometer-scale AlO$_{x}$-based MTJs in Ref.~\onlinecite{Sousa2004}, where scanning tunnel microscopy (STM) revealed regions of high current density within the tunnel barrier layer. Such hot spots were estimated to occupy about 5\% of the total barrier area ($A_{\textrm{cs}}$), which was enough to enhance the heating effect of current substantially. We model hot spots by reducing the effective area ($A^{\textrm{eff}}_{\textrm{TB}}$) of the tunnel barrier, which indeed increases the calculated heating efficiency and the final temperatures achieved for a given applied current [see Fig.~\ref{figure3}(b)]. From the comparison of $R(I)$ and $M(T)$ in Fig.~\ref{figure2} we estimate $A^{\textrm{eff}}_{\textrm{TB}} \approx 0.035 A_{\textrm{cs}}$, which correlates well with the earlier STM study. 

The simulated temperature profiles for $A^{\textrm{eff}}_{\textrm{TB}} \approx 0.035 A_{\textrm{cs}}$ show the expected, pronounced dependence on the current magnitude [Fig.~\ref{figure3}(a)]. The temperature in the stack peaks at the tunnel barrier (AlO$_x$), where the strongest inelastic electron-phonon relaxation takes place. The steep temperature profiles in the antiferromagnetic layers (IrMn) are due to the material's low heat conductivity, which has a positive effect of reducing heat dissipation into the Cu electrodes, thereby localizing the heating to the central part of the multilayer element. This also improves the temperature uniformity across the key spacer layer (Ni-Cu). 

The temperature of the Ni-Cu layer is a critical parameter as it induces the sequence of Curie and then magnetic switching in the structure. Figure~\ref{figure3}(a) shows that there is a significant temperature gradient across this layer. A reduction of this gradient by, for example, decreasing its thickness should make the thermo-magnetic switching sharper and more application friendly. Further optimization of the composition, morphology, and magnetic parameters of all the layers in the stack is undoubtedly possible. In particular, miniaturization and optimized heat focusing should greatly reduce the device operating power requirements. 

In conclusion, current-induced thermo-magnetic switching in a magnetic tunnel junction integrated into a vertical pillar is demonstrated. The effect is due to exchange coupling/decoupling of a soft ferromagnetic layer by a weakly ferromagnetic spacer layer, the Curie point of which is controlled by the heating effect of the transport current through the junction. The key aspects of the thermal layout of the device are analyzed and used to point out directions for optimizing the performance of the demonstrated device. 

Support from the Swedish Stiftelse Olle Engkvist Byggm{\"a}stare, the Swedish Research Council (VR Grant No. 2014-4548), the National Academy of Sciences of Ukraine (project 0115U00974), and the State Fund for Fundamental Research of Ukraine (project F76/63-2017) is gratefully acknowledged.

\bibliography{manuscript_library}
\end{document}